\documentclass[twoside]{IEEEtran}

\usepackage{caption}

\usepackage{subcaption}
\usepackage[utf8]{inputenc}

\usepackage{amssymb}
\usepackage{amsfonts}
\usepackage{amsmath}
\usepackage{graphicx}
\usepackage{indentfirst}
\usepackage{cite}

\usepackage{mathtools}

\usepackage{enumerate}
\usepackage{url}

\usepackage{cuted}

\usepackage[symbol]{footmisc}

\usepackage{tikz}
\definecolor{darkblue}{rgb}{0.15,0.35,0.55}
\definecolor{reddish}{rgb}{.8, 0.2, 0.2}
\usepackage[linktocpage=true]{hyperref}
\hypersetup{
colorlinks=true,
citecolor=darkblue,
linkcolor=reddish,
urlcolor=darkblue,
pdfauthor={},
pdftitle={},
pdfsubject={}
}

\long\def\ca#1\cb{} 

\newcommand{\ad}{^\dagger }

\newcommand{\becs}{\begin{cases}}
\newcommand{\bem}{\begin{matrix}}

\newcommand{\dya}[1]{|#1\rangle\langle#1|}
\newcommand{\dyad}[2]{|#1\rangle\langle#2|}

\newcommand{\encs}{\end{cases}}
\newcommand{\enm}{\end{matrix}}

\newcommand{\ket}[1]{|#1\rangle }

\newcommand{\mte}[2]{\langle#1|#2|#1\rangle }

\newcommand{\ot}{\otimes }

\newcommand{\ra}{\rightarrow }

\newcommand{\st}{\sqrt{2}}

\newcommand{\Tr}{{\rm Tr}}

\newcommand{\BC}{{\mathcal B}}
\newcommand{\CC}{{\mathcal C}}
\newcommand{\DC}{{\mathcal D}}
\newcommand{\EC}{{\mathcal E}}

\newcommand{\HC}{{\mathcal H}}
\newcommand{\IC}{{\mathcal I}}

\newcommand{\QC}{{\mathcal Q}}

\newcommand{\TC}{{\mathcal T}}

\newcommand{\rB}{\textbf{r}}

\newcommand{\al}{\alpha }
\newcommand{\bt}{\beta }

\newcommand{\dl}{\delta }
\newcommand{\Dl}{\Delta }
\newcommand{\ep}{\epsilon}

\newcommand{\zt}{\zeta }

\newcommand{\lm}{\lambda }

\newcommand{\sg}{\sigma }

\begin{document}
\title{Positivity and nonadditivity of quantum capacities using generalized
erasure channels}
\author{Vikesh~Siddhu, and~Robert~B.~Griffiths 
\thanks{
        Vikesh Siddhu was with the Department of Physics, Carnegie Mellon
        University, Pittsburgh, Pennsylvania 15213, USA. He is now with JILA,
        University of Colorado/NIST, Boulder, CO 80309, U.S.A.  (e-mail:
        vsiddhu@protonmail.com)
}
\thanks{
    Robert B. Griffiths is an emeritus Professor of Physics at Carnegie Mellon
    University in Pittsburgh, PA 15213. (email: rgrif@cmu.edu) 
}
\thanks{
    Copyright (c) 2014 IEEE. Personal use of this material is permitted.
    However, permission to use this material for any other purposes must be
    obtained from the IEEE by sending a request to pubs-permissions@ieee.org.
}
}
\date{5 Oct 2021}
\maketitle

\begin{abstract}

We consider various forms of a process, which we call {\em gluing}, for
    combining two or more complementary quantum channel pairs $(\BC,\CC)$ to
    form a composite. One type of gluing combines a perfect channel with a
    second channel to produce a \emph{generalized erasure channel} pair
    $(\BC_g,\CC_g)$.  We consider two cases in which the second channel is (i)
    an amplitude-damping, or (ii) a phase-damping qubit channel; (ii) is the
    \emph{dephrasure channel} of Leditzky et al. For both (i) and (ii),
    $(\BC_g,\CC_g)$ depends on the damping parameter $0\leq p\leq 1$ and a
    parameter $0 \leq \lm \leq 1$ that characterizes the gluing process. In
    both cases we study $Q^{(1)}(\BC_g)$ and $Q^{(1)}(\CC_g)$, where $Q^{(1)}$
    is the channel coherent information, and determine the regions in the
    $(p,\lm)$ plane where each is zero or positive, confirming previous results
    for (ii). A somewhat surprising result for which we lack any intuitive
    explanation is that $Q^{(1)}(\CC_g)$ is zero for $\lm \leq 1/2$ when $p=0$,
    but is strictly positive (though perhaps extremely small) for all values of
    $\lm> 0$ when $p$ is positive by even the smallest amount. In addition we
    study the nonadditivity of $Q^{(1)}(\BC_g)$ for two identical channels in
    parallel.  It occurs in a well-defined region of the $(p,\lm)$ plane in
    case (i). In case (ii) we have extended previous results for the dephrasure
    channel without, however, identifying the full range of $(p,\lm)$ values
    where nonadditivity occurs.  Again, an intuitive explanation is lacking.

\end{abstract}


\IEEEpeerreviewmaketitle

\section{Introduction}
\label{Sintro}

Understanding the capacity of a noisy quantum channel to transmit information
is a central and challenging problem in quantum information theory. In contrast
to the case of a classical channel one can define several capacities for a
quantum channel, among them the capacity to transmit classical
information~\cite{Holevo98,SchumacherWestmoreland97}, the private
capacity~\cite{Devetak05}, and---the subject of the present paper---the
\emph{quantum capacity}, a measure of its ability to transmit quantum
information. The asymptotic capacity $C$ of a classical channel was shown by
Shannon~\cite{Shannon48} to be equal to $C^{(1)}$, the {\em channel mutual
information} i.e., the mutual information between input and output, when
maximized over probability distributions of the input. An analog of the mutual
information for a quantum channel $\BC$ is the \emph{entropy bias} or {\em
coherent information} $\Dl(\BC,\rho)$~\cite{SchumacherNielsen96}, the
difference of the von Neumann entropies of the outputs of $\BC$ and its
\emph{complementary channel} $\CC$ for a given input density operator $\rho$.
Maximizing this over $\rho$ yields a non-negative real number, the
\emph{channel coherent information} (sometimes also called the
\emph{single-letter quantum capacity}) $Q^{(1)}(\BC)$.  (A quantum channel
$\BC$ of the kind considered here is always a member of a complementary pair of
channels $(\BC,\CC)$, $\CC$ the complement $\BC$ and vice versa, generated by a
single isometry as discussed in Sec.~\ref{Sprelim}.)

A significant difference between quantum and classical channels is that when
two classical channels are placed in parallel, the channel mutual information
$C^{(1)}$ of the combination is simply the sum of the individual channel mutual
informations, whereas in the quantum case when channel $\BC$ is placed in
parallel with $\BC'$ one has only an inequality:
\begin{equation}
    Q^{(1)}(\BC \ot \BC') \geq Q^{(1)}(\BC)  + Q^{(1)}(\BC').
    \label{eqm1}
\end{equation}
This inequality can be strict, i.e., $Q^{(1)}$ can be
\emph{nonadditive}~\cite{DiVincenzoShorEA98}. Nonadditivity makes it
difficult to calculate the asymptotic quantum capacity $Q(\BC)$ of a channel
$\BC$, the limit as $n\ra\infty$ of 
$Q^{(1)}(\BC^{\ot n})/n$~\cite{Lloyd97, Shor02a, Devetak05}. 
The asymptotic capacity $C$ of two classical channels placed in parallel
is simply the sum of the capacities of the individual channels. By contrast,
due to nonadditivity~\eqref{eqm1} the asymptotic capacity $Q(\BC \ot \BC')$ of two
quantum channels $\BC$ and $\BC'$ used in parallel may be greater than
$Q(\BC) + Q(\BC')$~\cite{SmithYard08, BrandaoOppenheimEA12}, 
which implies that the asymptotic $Q$, unlike its classical counterpart $C$,
does not completely capture a channel's ability to transmit quantum
information.  The mathematical or physical principles behind nonadditivity are
at present not well understood. Simple examples of nonadditivity are hard to
construct. One source of difficulty is finding the global maximum of the function
$\Dl(\BC,\rho)$ which in general is not a concave function of $\rho$.

If both $\BC$ and $\BC'$ are channels that are either degradable or
antidegradable~(see comments at the end of Sec.~\ref{Sprelim}) it is known that
\eqref{eqm1} is an equality, and therefore $Q(\BC)=Q^{(1)}(\BC)$, and
$Q(\BC\ot\BC') = Q(\BC)+Q(\BC')$~\cite{DevetakShor05, LeditzkyDattaEA18}.  For an antidegradable
channel $Q^{(1)}(\BC) = Q(\BC) =0$, and the same is true for
entanglement-binding channels~\cite{HorodeckiHorodeckiEA00}. But apart from special
cases such as these it is in general not easy to determine whether $Q^{(1)}$ or
$Q$ is positive or zero~\cite{CubittElkoussEA15,SmithSmolin12}.

For the case of two identical channels in parallel, $\BC'=\BC$, a simple
example of nonadditivity has recently been constructed by Leditzky et
al.~\cite{LeditzkyLeungEA18} using what they call the {\em dephrasure channel}.
To show nonadditivity they first find $Q^{(1)}(\BC)$, and then make a guess
or {\em ansatz} $\hat \rho_2$ for a bipartite input density operator for which
$\Dl(\BC^{\ot 2}, \hat \rho_2)$, a lower bound for $Q^{(1)}(\BC^{\ot 2})$, is
larger than $2 Q^{(1)}(\BC)$. The ansatz approach can be extended to $n$
identical channels in parallel in an obvious way to look for cases where
$\Dl(\BC^{\ot n},\hat\rho_n)$~(and hence $Q^{(1)}(\BC^{\ot n})$) exceeds
$nQ^{(1)}(\BC)$. This approach has been successfully applied for $n \geq 5$ to
the qubit depolarizing channel~\cite{DiVincenzoShorEA98} where $Q^{(1)}$ is
known, and to other qubit Pauli channels~\cite{SmithSmolin07, FernWhaley08,
BauschLeditzky19} where $Q^{(1)}$ is believed to be known.

Our exploration of some of these issues begins with a general procedure for
combining several quantum channels to form a new channel through a process we
call \emph{gluing}. It differs from the familiar procedures of placing channels
in parallel or series, and it puts together in a single overall structure
concepts such as subchannels~\cite{SiddhuGriffiths16}, direct
sums~\cite{FukudaWolf07}, and convex sums of channels~(see Sec.~2.2
in~\cite{Watrous18}). A particular type of gluing results in what we call a
\emph{block diagonal} channel pair, an instance of which is the much-studied
and well-understood \emph{erasure channel}~\cite{BennettDiVincenzoEA97} with
erasure probability $0\leq \lm \leq 1$, whose complement is also an erasure
channel. The erasure channel can be regarded as the result of gluing together
two perfect channels as discussed in Sec.~\ref{SglnEra}. When one of the
perfect channel pairs is replaced by an arbitrary complementary channel pair
$(\BC_1, \CC_1)$ the result is a \emph{generalized erasure channel pair}
$(\BC_g, \CC_g)$. The $\BC_g$ channel can be viewed as a concatenation of
$\BC_1$ with an erasure channel, and $\CC_g$ as an ``incomplete erasure''
channel.

We study two cases of such generalized erasure channel pairs. In the first,
$\BC_1$ is a qubit-to-qubit amplitude damping channel, as is its complement
$\CC_1$. In the second, $\BC_1$ is a qubit-to-qubit phase-damping channel whose
complement is a measure-and-prepare channel; here $\BC_g$ is the dephrasure
channel. In both cases the qubit channel pair $(\BC_1, \CC_1)$ depends on a
parameter $0 \leq p \leq 1$, and thus $(\BC_g,\CC_g)$ depends on two
parameters, $p$ and the erasure probability $\lm$. For all values of these
parameters we compute $Q^{(1)}(\BC_g)$ and $Q^{(1)}(\CC_g)$ by performing
a global optimization and find the $(p,\lm)$ values for which they are
positive. The dependence of $Q^{(1)}(\CC_g)$ on these parameters is rather
surprising---see Fig.~\ref{Q1C} and the accompanying discussion---and worth
further study.

In both the amplitude and phase damping cases we find nonadditivity, a strict
inequality in \eqref{eqm1}, when both $\BC$ and $\BC'$ are $\BC_g$. Our results
in the amplitude damping case indicate that nonadditivity occurs over a
well-defined region in the space $(p,\lm)$ of parameters, as shown in 
Fig.~\ref{FigY}.
For the phase-damping case, where $\BC_g$ is the dephrasure channel, our
numerical results confirm and also extend the region of nonadditivity
identified in~\cite{LeditzkyLeungEA18}, but without finding its precise boundaries. In
addition we have carried out a limited exploration of higher-order
nonadditivity by using various ansatzes, but without finding anything very
interesting.

The remainder of this paper is structured as follows. Section~\ref{Sprelim}
contains preliminary definitions and notation: in particular our use of
isometries to construct a channel pair, and the use of projective
decompositions of the identity (PDIs) to identify orthogonal subspaces.
Definitions of the entropy bias $\Dl(\BC,\rho)$ and the channel coherent
information $Q^{(1)}(\BC)$ of a channel $\BC$, and (anti)degradable channels
are also found in this section.
Various gluing procedures for combining two or more channels are discussed at
some length in Sec.~\ref{Sgluing}. The particular procedure that yields a block
diagonal channel pair~(see eq. \eqref{blockDiag}) is employed in
Sec.~\ref{SglnEra} to define a generalized erasure channel pair. The
amplitude-damping case is discussed in Sec.~\ref{SAmpDamp}, and the
phase-damping~(dephrasure) case in Sec.~\ref{Sdephasing}. The surprising
positivity of $Q^{(1)}$ in the ``incomplete erasure'' situation found in both
cases is discussed in Sec.~\ref{SEra}. A concluding Sec.~\ref{Scon}
contains a summary of our results and an indication of some unsolved problems
that deserve further study. It is followed by two appendices devoted to some
technical issues and details.

\section{Preliminaries}
\label{Sprelim}

A quantum channel (completely positive trace preserving map) can always 
be constructed using an isometry $J$ 
\begin{equation}
   J: \HC_a \mapsto \HC_b \ot \HC_c; \quad 
    J^{\dag}J = I_a,
    \label{isoCond}
\end{equation}
mapping the Hilbert space $\HC_a$ representing the channel's input to a
subspace of $\HC_b \ot \HC_c$, where $\HC_b$ and $\HC_c$ represent the direct
and complementary channel outputs. The isometry preserves inner products, and
this is ensured by the condition $J^{\dag}J = I_a$, where $I_a$ is the identity
operator on $\HC_a$. We assume that the dimensions $d_a$, $d_b$ and $d_c$ of
the three Hilbert spaces are finite and satisfy $d_a \leq d_b d_c$, but are
otherwise unrestricted. The isometry results in a pair of quantum channels
with superoperators
\begin{equation}
    \BC(A) = \Tr_c(J A J^{\dag}),\quad
\CC(A) = \Tr_b(J A J^{\dag}),
    \label{chanDef}
\end{equation}
that map $\hat \HC_a$, the space of operators on $\HC_a$, to the operator
spaces $\hat \HC_b$ and $\hat \HC_c$, respectively. Given the superoperator
$\BC$, the corresponding isometry $J$, and thus the superoperator $\CC$, is
uniquely determined up to a unitary acting on $\HC_c$
\footnote{This assumes the complementary output space $\HC_c$ is as small as
possible, which is to say it is the support of $\CC(I_a)$. If one allows
$\HC_c$ to have higher dimension than the support, ``unitary'' must be replaced
with ``partial isometry'', see Sec.~5.2 in \cite{Wilde17}.}.
Likewise, $\CC$ determines $\BC$ up to a unitary on $\HC_b$. 
One refers to $\CC$ as the \emph{complement} of
$\BC$, or $\BC$ as the complement of $\CC$, and the two together as a 
\emph{complementary pair} of channels. 

We shall be concerned with orthogonal subspaces of $\HC_a$, $\HC_b$ and $\HC_c$,
and it is convenient to represent the subspaces using a 
\emph{projective decomposition of the identity}~(PDI), a collection of 
mutually orthogonal projectors that
sum to the identity. Thus a PDI $\{P_j\}$ of $\HC_a$ is a set of projectors
such that
\begin{equation}
    P_j = P_j^{\dag} = P_j^2, \quad
    P_jP_j = \dl_{ij}P_j, \quad \text{and} \quad
    \sum_j P_j = I_{a}.
    \label{PDI}
\end{equation}
Using it one can define the $j'$th subspace $\HC_{aj}$ of $\HC_a$,
as
\begin{equation}
    \HC_{aj} = P_j \HC_a;
    \label{subspacePDI}
\end{equation}
that is to say, the collection of all $\ket{\psi}$ such that $P_j\ket{\psi} =
\ket{\psi}$, which means they are orthogonal to any $\ket{\phi}$ in 
$\HC_{ak}$ with $k\neq j$. Thus $\HC_a$ is a direct sum,
\begin{equation}
    \HC_a = \bigoplus_j \HC_{aj}, 
    \label{dirSum}
\end{equation}
of these orthogonal subspaces. Similarly, a PDI $\{Q_j\}$ can be used to
partition $\HC_b$ into subspaces $\HC_{bj}=Q_j\HC_b$, and  $\{R_j\}$
to partition $\HC_c$ into $\HC_{cj} = R_j\HC_c$.

The \textit{coherent information} or \textit{entropy bias} of a channel
$\BC$ with complement $\CC$ for an input density operator $\rho$ in $\hat\HC_a$
is
\begin{equation}
\Dl(\BC,\rho) = S(\BC(\rho)) - S(\CC(\rho)),
\label{entBias}    
\end{equation}
where $S(\rho) = - \Tr(\rho \log_2 \rho)$ is the von-Neumann entropy of
$\rho$~(in base $2$). 
Since the complement $\CC$ is determined by $\BC$ up to a unitary~(see the
comments following eq.~\eqref{chanDef}) which does not change the von-Neumann
entropy, the entropy bias depends only on $\BC$.
Since the entropy bias is the difference of two entropy functions, each of
which is concave in $\rho$, the bias itself need not be concave or convex in
$\rho$. The \textit{channel coherent information} sometimes known (slightly
confusingly) as the \textit{single-letter quantum capacity}, for each channel
in the $(\BC,\CC)$ pair is given by
\begin{equation}
    Q^{(1)}(\BC) = \max_{\rho} \Dl(\BC,\rho), \quad  
    Q^{(1)}(\CC) = -\min_{\rho} \Dl(\BC,\rho).
    \label{q1Def}
\end{equation}
The channel $\BC$ is said to be \emph{degradable} and $\CC$
\emph{antidegradable} if there exists a quantum channel $\DC$ such that
$\CC=\DC\circ\BC$, i.e., if the output of $\BC$ is made the input of $\DC$, the
result is $\CC$. Various properties of such channel pairs are discussed
in~\cite{DevetakShor05, Holevo08, CubittRuskaiEA08}.  Of particular relevance
in what follows is the fact that $Q(\CC) = Q^{(1)}(\CC) =0$ for an
antidegradable channel $\CC$; and the entropy bias $\Dl(\BC,\rho)$ of a
degradable channel $\BC$ is a concave function of
$\rho$~\cite{YardHaydenEA08}, making it relatively easy to compute its maximum
$Q^{(1)}(\BC)=Q(\BC)$.

\section{Glued Isometries and Channels}
\label{Sgluing}

There are various ways of combining quantum channels and their corresponding
isometries. A \emph{concatenation} of two channels in which the output of the
first becomes the input of the second corresponds to the concatenation of the
two isometries. When two channels or channel pairs are placed in parallel, the
input space of the combination is the tensor product of the two input spaces,
likewise the direct and complementary output spaces are the corresponding tensor
products, and the isometry for the combination is the tensor product of the
individual isometries. But in addition isometries can be combined in such a way
that one or more of the input, direct output and complementary output spaces
are \textit{subspaces} of larger Hilbert spaces, a process which we refer to as
\textit{gluing}. The idea will become clear from the following examples.

Consider a collection of isometries 
\begin{equation}
 J_j: \HC_{aj} \mapsto \HC_{bj}\ot\HC_{cj}
\label{isoms}
\end{equation}
in the notation of Sec.~\ref{Sprelim}, where the $\HC_{aj}$ are either
distinct orthogonal subspaces of $\HC_a$, or all equal to $\HC_a$, and the same
for the $\HC_{bj}$ and $\HC_{cj}$; see \eqref{subspacePDI} and \eqref{dirSum}.
We shall, in what follows, assume the convenient, but not absolutely necessary,
condition:
\begin{equation}
     J_j^{\dag}J^{}_k = 0 \; \text{for} \; j \neq k.
    \label{coll}
\end{equation}
Finally, the overall isometry $J : \HC_a \mapsto \HC_b \ot \HC_c$, obtained
from gluing the collection of isometries in \eqref{isoms}, is given by a sum
\begin{equation}
    J := \sum_j \nu_j J_j,
    \label{glueEq}
\end{equation}
where the $\nu_j$ are positive numbers. The condition $J\ad J=I_a$ for
$J$ to be an isometry is then
\begin{equation}
    \sum_j \nu_j^2 J_j^{\dag} J^{}_j = I_a.
    \label{sumCond}
\end{equation}
The isometry $J$ in \eqref{glueEq} defines a pair of
channels $\BC$ and $\CC$ through \eqref{chanDef}. 

Perhaps the simplest example of gluing is when each $J_j$ is simply
$J$ applied to the subspace $\HC_{aj}=P_j\HC_a$, while $\HC_{bj}=\HC_b$
and $\HC_{cj}=\HC_c$, independent of $j$, and thus
\begin{equation}
    J_j =J P_j,\quad  
    J_j^{\dag} J^{}_j = P_j J^{\dag} J P_j = P_j I_a P_j = P_j,
    \label{isoDefA}
\end{equation}
where the projector $P_j$ is the identity operator on $\HC_{aj}$.
Then with every $\nu_j=1$ in \eqref{glueEq} one has
\begin{equation}
    J = \sum_j J_j = \sum_j J P_j = J I_a.
    \label{glueA}
\end{equation}
The corresponding subchannels $\BC_j$ and $\CC_j$ are
given by the expressions
\begin{equation}
    \BC_j(A) = \BC(P_j A P_j),\quad
    \CC_j(A) = \CC(P_j A P_j),
    \label{subChan}
\end{equation}
in terms of the superoperators $\BC$ and $\CC$ for the full channel and its
complement. In general $\BC(A)$ will not equal $\sum_j \BC_j(A)$, because the
latter maps all ``off-diagonal'' parts, $P_j A P_k$ for $j\neq k$, of the
operator $A$ to zero; similarly, $\CC(A)$ will in general not be the sum of the
$\CC_j(A)$.

Another example of gluing arises given a collection of isometries
$J_j:\HC_a \mapsto \HC_b\ot\HC_{cj}$; that is, the direct channels have the
same input and output spaces $\HC_a$ and $\HC_b$, whereas the complementary
channels map to orthogonal subspaces $\HC_{cj}=R_j\HC_c$ of $\HC_c$. If we
write \eqref{glueEq} in the form
\begin{equation}
    J = \sum_j \sqrt{p_j}\, J_j,
    \label{glueC}
\end{equation}
i.e., $\nu_j=\sqrt{p_j}$, where the $p_j>0$ are any set of probabilities that
sum to $1$, then
\begin{equation}
 J_j = R_j J/\sqrt{p_j},
\label{glueC2}
\end{equation}
and it is easily checked that
\eqref{coll} and \eqref{sumCond} are satisfied. It
is straightforward to show that the $\BC$ channel resulting from this gluing of
the $\HC_{cj}$ spaces is given by
\begin{equation}
    \BC(A) = \sum_j p_j \BC_j(A),
    \label{cvxComb}
\end{equation}
thus a weighted sum or \emph{convex
combination} of the $\BC_j$ channels corresponding to the different $J_j$
isometries. Furthermore, \emph{any} convex combination $\sum_j p_j\BC_j$ of
channels with a common input space $\HC_a$ and output space $\HC_b$ can be
constructed in this manner by gluing together the different $\HC_{cj}$ output
spaces of the complementary channels. But in general there is no simple
relationship between the complementary channel $\CC$ and the different $\CC_j$.

One can combine the two previous examples and glue both the input spaces
$\HC_{aj}$ and the complementary output spaces $\HC_{cj}$ of isometries
$J_j : \HC_{aj} \ra \HC_{b} \ot \HC_{cj}$, which have a common direct output
space $\HC_b$. With $\nu_j=1$ in
\eqref{glueEq} one has
\begin{equation}
 J = \sum_j J_j,\quad J_j = R_j J P_j.
\label{glueAC}
\end{equation}
Again \eqref{coll} is obviously satisfied. A simple calculation shows that
\begin{equation}
    \BC(A) = \sum_j\BC(P_jAP_j) = \sum_j \BC_j(A),
\label{glueAC2}
\end{equation}
and that $\BC_j$ and $\CC_j$ satisfy \eqref{subChan}. But now $\BC$ (as well as
$\BC_j$ and $\CC_j)$ maps ``off diagonal'' parts, $P_j A P_k$ for $j\neq k$ 
of the operator $A$ to zero. There is in general no connection 
between $\CC$ and the $\CC_j$ analogous to \eqref{glueAC2}.

Of particular interest for what follows later is a \emph{block diagonal}
channel pair obtained from gluing both the direct and complementary outputs of
the isometries $J_j:\HC_a\mapsto \HC_{bj}\ot\HC_{cj}$, with $\HC_{bj} = Q_j
\HC_b$ and $\HC_{cj} = R_j\HC_c$. As in \eqref{glueC} one writes
\begin{equation}
    J = \sum_j\sqrt{p_j}\, J_j,\quad J_j = (Q_j \ot R_j ) J/\sqrt{p_j}.
\label{glueBC}
\end{equation}
 It is then straightforward to show that 
\begin{equation}
    Q_j\BC(A) Q_k = \dl_{jk} p_j \BC_j(A),\quad 
    R_j\CC(A) R_k = \dl_{jk} p_j \CC_j(A),
    \label{BDeqn}
\end{equation} 
and as a consequence, 
\begin{equation}
    \BC(A) = \bigoplus_j  p_j \BC_j(A), \quad
    \CC(A) = \bigoplus_j  p_j \CC_j(A).
    \label{blockDiag}
\end{equation}
Instead of $\bigoplus$, one could have used $\sum$ in \eqref{blockDiag} to
indicate that both $\BC$ and $\CC$ are convex sums of $\{\BC_i\}$ and
$\{\CC_i\}$ respectively; however, using $\bigoplus$ rather than $\sum$ serves
to emphasize that the output spaces of the $\{\BC_i\}$ are mutually orthogonal,
as are the output spaces of the $\{\CC_i\}$. Thus the outputs of each channel
in the $(\BC, \CC)$ pair are in separate, mutually orthogonal blocks, whence
our name `block diagonal' channel pair. In addition, from
eq.~\eqref{glueBC} it follows that the $\BC$ and $\CC$ blocks are
\emph{correlated}: if in a particular run the output of the $\BC$ channel falls
in a particular block $Q_j$ (as could be determined by a suitable measurement),
the $\CC$ channel output will be in the corresponding block $R_j$. This means
the entropy of the $\BC$ output is given by
\begin{equation}
    S\big(\BC(\rho)\big) = \sum_j p_j S\big(\BC_j(\rho)\big) + h(p),
    \label{concEntB}
\end{equation}
where $h(p) = -\sum_j p_j \log_2 p_j$ is the Shannon entropy of the probability
distribution $\{p_i\}$~(in base 2). There is an analogous expression for the
entropy of the $\CC$ output. Thus the output entropy in each case is the
weighted sum of the output entropies of the individual channels plus a
"classical" term $h(p)$. This classical term cancels when one computes the
entropy bias, \eqref{entBias}, which is given by
\begin{equation}
    \Dl(\BC,\rho) = \sum_j p_j \Dl(\BC_j,\rho).
    \label{concEntBias}
\end{equation}
These considerations suggest a simple physical picture:  the channel $\BC$ can
be obtained by randomly applying with probability $p_i$ the channel 
$\BC_i$ to the input $\hat\HC_a$, with the output going to
$\hat\HC_{bi}$. Thus $\BC$ is a convex combination of the $\BC_i$ if one
regards each of these as a map into the full operator space $\hat\HC_b$. 
A similar comment applies to $\CC$ as a convex combination of the $\CC_i$.
The structure of these block diagonal channel pairs is exemplified by the
specific examples in Sec.~\ref{SglnEra}.

It is possible to glue the inputs and both the direct and complementary outputs
in a construction called the \emph{direct sum} of
channels. Let the corresponding isometries be
$J_j:\HC_{aj} \mapsto \HC_{bj}\ot\HC_{cj}$, with $\HC_{aj} = P_j\HC_a$,
$\HC_{bj} = Q_j\HC_b$, $\HC_{cj} = R_j\HC_c$. Thus the corresponding channels
are completely independent of each other, with distinct input and output
spaces. With $J$ the sum of these isometries one has:
\begin{equation}
   J = \sum_j J_j; \quad J_j = (Q_j \ot R_j) J P_j.
    \label{gIsoDirSum}
\end{equation}
A straightforward calculation shows that $J$ gives rise to 
channels 
\begin{align}
    \BC(A) = \bigoplus_j \BC_j(P_j A P_j), \quad
    \CC(A) = \bigoplus_j \CC_j(P_j A P_j).
    \label{dirSumChan}
\end{align}
Once again there is a block-diagonal structure with correlated blocks, but 
the physical picture
is a bit different from \eqref{blockDiag}. The channel $\BC$ acts on a 
density operator $\rho$ by 
projecting it to the sub-space $\HC_{aj}$
with probability $\Tr(P_j\rho)$~(thus the ``off-diagonal'' $P_j \rho P_k$
parts of $\rho$ for $j\neq k$ always map to zero), and then 
applying the channel $\BC_j$. An analogous interpretation holds for $\CC$.
This direct sum construction of a channel $\BC$ has been studied
in~\cite{FukudaWolf07} where it was used in simplifying the nonadditivity
conjecture of the Holevo capacity of a quantum channel. 
The gluing picture reveals that $\CC$, the complement of $\BC$, is also
a direct sum, and the two channels $\BC$ and $\CC$ have correlated 
blocks.

If an isometry $J$ has been produced by gluing other isometries together in the
manner indicated above, one can recover the {constituents by a process of
\emph{slicing} $J$, in which projectors corresponding to the different PDIs are
placed to the left and right of $J$. For any map $J$ from $\HC_a$ to $\HC_b\ot
\HC_c$, not necessarily an isometry, one can define a collection of operators
\begin{equation}
    K_{jkl} := (Q_k \ot R_l) J P_j. 
\label{eqn22}
\end{equation}
using PDIs $\{P_j\}$, $\{Q_k\}$, and $\{R_l\}$, which need not have the same
number of projectors, on $\HC_a$, $\HC_b$, and $\HC_c$, respectively. It is
obvious that $J$ is the sum of all of the $K_{jkl}$, but even if $J$ is an
isometry, the individual $K_{jkl}$ will in general \emph{not} be isometries or
proportional to isometries; that is, $K_{jkl}\ad K^{}_{jkl}$ will not be
proportional to $P_j$. Only for special choices of the isometry $J$ and the
PDIs, as in the examples considered above, will the operators resulting from
slicing be proportional to isometries.

\section{Generalized Erasure Channel Pair}
\label{SglnEra}

An \emph{erasure channel} pair is an example of a block diagonal channel
with two blocks, where the isometries in \eqref{glueBC} are given by
\begin{equation}
    J_1 \ket{\psi}_{a} = \ket{\psi}_{b1} \ket{e}_{c1}, \quad
    J_2 \ket{\psi}_{a} = \ket{f}_{b2}\ket{\psi}_{c2}.
    \label{erasIso}
\end{equation}
Here $\HC_{b1}$ and $\HC_{c2}$ are isomorphic to $\HC_{a}$, whereas $\HC_{c1}$
and $\HC_{b2}$ are one-dimensional Hilbert spaces spanned by the
normalized kets $\ket{e}_{c1}$ and $\ket{f}_{b2}$, respectively. By
definition, $\HC_{b1}$ and $\HC_{b2}$ are orthogonal subspaces of $\HC_b =
\HC_{b1} \oplus \HC_{b2}$, and $\HC_{c1}$ and $\HC_{c2}$ are orthogonal
subspaces of $\HC_c = \HC_{c1} \oplus \HC_{c2}$
The isometry $J_1$ generates a {\em perfect channel pair} $(\IC, \TC)$, where
the identity channel $\IC$ maps any operator $A$ in $\hat \HC_a$ to the same
operator $A$ in $\hat \HC_{b1}$, while the trace channel $\TC$ maps $A$ to
$\Tr(A) [e]$ in $\hat \HC_{c1}$. Here and later we use the abbreviation
$[\psi]=\dya{\psi}$ for the projector corresponding to a normalized ket
$\ket{\psi}$. In the same way $J_2$ generates the perfect channel pair $(\TC,
\IC)$ mapping $\hat \HC_a$ to $\hat \HC_{b2}$ and $\hat \HC_{c2}$,
respectively.
Gluing these perfect channel pairs together using $p_1=1-\lm$ and $p_2=\lm$ in
\eqref{glueBC} results in the channel pair
\begin{align}
    \BC_e(A) &= \EC^{\lm}(A) = (1-\lm)A \bigoplus \lm \Tr(A) [f], 
    \label{er1} \\
    \CC_e(A) &= \EC^{1-\lm}(A) = (1-\lm)\Tr(A) [e] \bigoplus \lm A.
    \label{er2}
\end{align}
The channel $\EC^{\lm}$ is the \emph{erasure channel} with
erasure probability $\lm$, and $\EC^{1-\lm}$ is its complement. The subspaces
on the left and right side of $\bigoplus$ can be interchanged; the order used in
\eqref{er1} and \eqref{er2} reflects the correlations discussed following
\eqref{blockDiag}: if $A$ occurs in the output of the direct channel, $[e]$
will be present in the complementary output.

It is easily shown that the entropy bias, \eqref{entBias}, of $\BC_e=\EC^\lm$
takes the form
\begin{equation}
\Dl(\EC^\lm,\rho) =(1-2\lm)S(\rho).
\label{bbias}
\end{equation}
Its maximum for $\lm \geq 1/2$ is zero, and $(1-2\lm)\log_2 d_a$ for $\lm \leq
1/2$, since the maximum value $\log_2 d_a$ of $S(\rho)$ is achieved when $\rho$
is proportional to the identity operator. If $\EC^\lm$ for $\lm \leq 1/2$ is
followed by $\EC^\mu$ with $\mu = (1 - 2 \lm)/( 1 - \lm)$, the resulting
channel $\EC^\mu\circ\EC^\lm=\EC^{1-\lm}$ is the complement of $\EC^\lm$. Hence
for $\lm\leq 1/2$, $\EC^\lm$ is degradable, while for $\lm \geq 1/2$ it is
antidegradable. Consequently $Q$ and $Q^{(1)}$ are identical for an erasure
channel, and 
\begin{align}
 Q^{(1)}(\BC_e) =Q(\BC_e) &= Q(\EC^\lm) = \max\{1 - 2\lm,0\} \log_2 d_a,
\notag\\
 Q^{(1)}(\CC_e) =Q(\CC_e) &= Q(\EC^{1-\lm}) = 
\max\{2\lm-1,0\} \log_2 d_a.
\label{capEq}
\end{align}

We define the \textit{generalized erasure channel} pair as one in which $J_2$
is the same as in \eqref{erasIso} and corresponds to a perfect channel, but
$J_1$ is replaced by \emph{any} isometry from $\HC_a$ to
$\HC_{b1} \ot \HC_{c1}$, where the dimensions of $\HC_{b1}$ and $\HC_{c1}$ are
arbitrary (except that the product cannot be less than the dimension of
$\HC_a)$. The result is a channel pair
\begin{align}
    \BC_g(A) &= (1 - \lm) \BC_1(A) \bigoplus \lm \Tr(A) [f], \label{glnEra1}\\
    \CC_g(A) &= (1 - \lm) \CC_1(A) \bigoplus \lm A,
    \label{glnEra2}
\end{align}
where $(\BC_1,\CC_1)$ is the channel pair generated by $J_1$. 
The form of $\BC_g$ in \eqref{glnEra1} means that it either erases its input
with probability $\lm$, or else sends it through $\BC_1$ into the output
subpace $\HC_{b1}$ of $\HC_b$. Similarly, $\CC_g$ with probability $\lm$ sends
its input unchanged to $\HC_{c2}$, or else sends it through $\CC_1$ to
$\HC_{c1}$. When $\CC_1$ is the trace channel $\TC$ that completely
erases its input, $\CC_g$ is an erasure channel $\CC_e$ with erasure
probability $1-\lm$. But in general $\CC_1$ need not erase completely, so we
call $\CC_g$ an ``incomplete erasure'' channel.

The $\BC_g$ channel can be obtained by concatenating $\BC_1$ either with a
preceding erasure channel $\EC^{\lm}$, defined in~\eqref{er1}, or one that
follows it, $\tilde \EC^{\lm}$:
\begin{equation}
  \BC_g(A) = \tilde \BC_1 \circ \EC^{\lm}(A) =\tilde \EC^{\lm} \circ \BC_1(A).
\label{reversal}
\end{equation}
Here the superoperator $\tilde \EC^{\lm}$ is an erasure channel whose input
space is identical to the output space of $\BC_1$, which need not have the same
dimension as its input space. The operator $\tilde\BC_1$ is the same as $\BC_1$
except that when it is applied to $[e]$ in \eqref{er1} $\tilde\BC_1$ maps $[e]$
to the corresponding $[e]$ in \eqref{glnEra1}, and maps any ``off-diagonal''
ket $\dyad{e}{\al}$ or $\dyad{\al}{e}$, $\ket{\al}$ any element of $\HC_a$, to
zero. Given these definitions it is straightforward to check the validity of
\eqref{reversal}.  There is no analog of \eqref{reversal} for $\CC_g(A)$.
Since the concatenation of a channel with an antidegradable channel always 
results in an antidegradable channel~(see App.~\ref{Aad} for a simple proof),
$\BC_g$ is antidegradable when either $\BC_1$ or $\EC^{\lm}$ is 
antidegradable (the latter happens when $\lm \geq 1/2$).

As a consequence of data-processing~\cite{SchumacherNielsen96}, a channel
obtained by concatenating two channels has a smaller channel coherent
information and hence smaller quantum capacity than either of the channels that
are being concatenated~\cite{SmithSmolin08, KhatriSharmaEA20}.
Thus, from \eqref{reversal} it follows that 
$Q^{(1)}(\BC_g) \leq \min \{Q^{(1)}(\BC_1),Q^{(1)}(\EC^{\lm}) \}$
and $Q(\BC_g) \leq \min \{Q(\BC_1),Q(\EC^{\lm}) \}$. 
The channel $\CC_e$ in \eqref{er2} can be obtained by concatenating the output of 
$\CC_g$ with a channel that traces out operators on $\HC_{c1}$ to a fixed pure
state $[f]$ and does nothing to $\HC_{c2}$, thus
$Q^{(1)}(\CC_e) \leq \min \{ Q^{(1)}(\CC_g), \log d_{a}\}$ and
$Q(\CC_e) \leq \min \{ Q(\CC_g), \log d_{a}\}$.

\section{Applications}
\label{SApp}

\subsection{Generalized Erasure using Qubit Amplitude Damping Channel}
\label{SAmpDamp}

The isometry $J_1:\HC_a \ra \HC_{b1}\ot\HC_{c1}$ defined by
\begin{align}
    J_1 \ket{0}_a &= \ket{0}_{b1}\ket{1}_{c1}, \nonumber \\ 
    J_1 \ket{1}_a &= \sqrt{1 - p}\, \ket{1}_{b1}\ket{1}_{c1} + 
                    \sqrt{p}\, \ket{0}_{b1}\ket{0}_{c1},
    \label{ampIso}
\end{align}
with $0\leq p \leq 1$, and $\ket{0}$ and $\ket{1}$ are the usual orthonormal
basis kets for a qubit, defines a channel pair $(\BC_1,\CC_1)$ in which $\BC_1$
is an \emph{amplitude-damping channel} with $p$ the probability that the input
state $[1]_a$ decays to the output state $[0]_{b1}$. Similarly, $\CC_1$ is an
amplitude damping channel with decay rate $(1-p)$ if one interchanges
$\ket{0}_{c1}$ and $\ket{1}_{c1}$ in \eqref{ampIso}. The Bloch vector
parametrization for a qubit density operator,
\begin{equation}
     \rho(\rB) = \frac{1}{2}(I + \rB.\vec{\sg})
     := \frac{1}{2}(I + x \sg_x + y \sg_y + z \sg_z),
    \label{qBitBloch}
\end{equation}
where $I$ is the identity and $(\sg_x,\sg_y,\sg_z)$ the three Pauli matrices,
provides a convenient way to represent $\BC_1$ and $\CC_1$ as maps carrying
$\rB$ to Bloch vectors
\begin{align}
    \nonumber
    \rB_b &= (\sqrt{1-p}\;x, \sqrt{1-p}\;y, (1-p)z + p)
    \quad \text{and} \\ 
    \rB_c &= (\sqrt{p}\;x, -\sqrt{p}\;y, p - pz -1),
    \label{qBitIso}
\end{align}
respectively. See Fig.~\ref{FigX}(a) for a convenient way to visualize
this channel pair. 

\begin{figure*}[ht]
    \centering
    \includegraphics[scale=1]{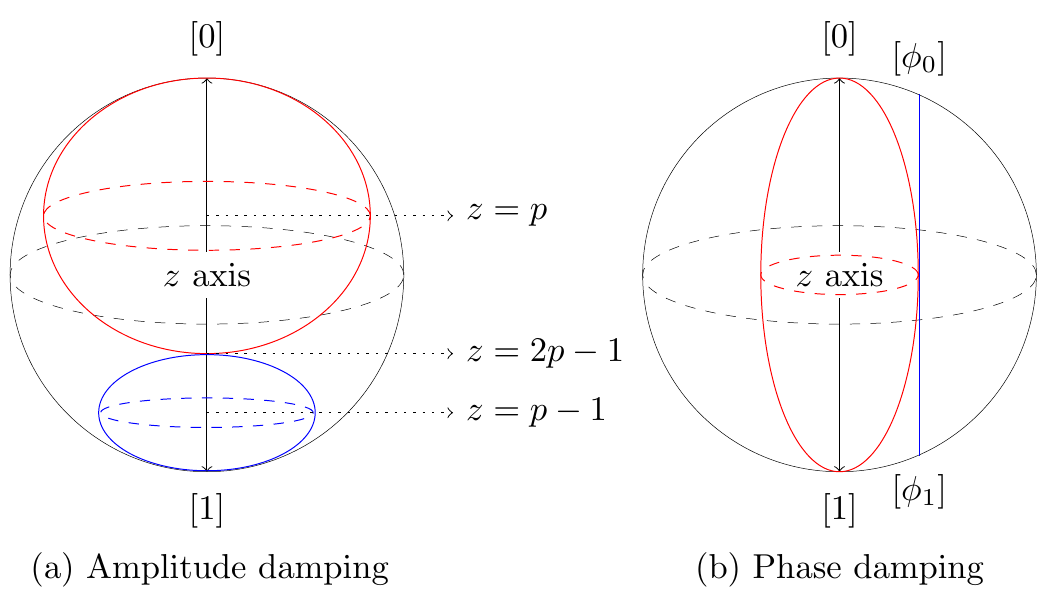}
    \caption{Enclosed inside a Bloch sphere~(in black): ellipsoids representing
    the loci of Bloch vectors $\rB_b$~(in red) and $\rB_c$~(in blue) for
    $p=0.3$. Square brackets indicate projectors, e.g., $[1]$ projects on
    $\ket{1}$.
 (a) Amplitude damping:  
     $\rB_b$ and $\rB_c$  defined in 
    \eqref{qBitIso}.
 (b) Phase damping: $\rB_b$ and $\rB_c$  defined in
    \eqref{dephDef}.}
    \label{FigX}
\end{figure*}

Let $(\BC_g,\CC_g)$   be the generalized channel pair resulting from
inserting $\BC_1$ and $\CC_1$ in \eqref{glnEra1} and \eqref{glnEra2}.  
The entropy bias $\Dl(\BC_g, \rho(\rB))$ of $\BC_g$ at $\rho(\rB)$ is a
real-valued function of $\rB = (x,y,z)$, whose maximum and minimum over $\rB$
with $|\rB| \leq 1$ gives $Q^{(1)}(\BC_g)$ and $Q^{(1)}(\CC_g)$,
respectively; see \eqref{q1Def}. Finding these extrema is simplified by
the fact that the rotational symmetry of $\Dl(\BC_g, \rho(\rB))$ about the $z$
axis---see Fig.~\ref{FigX}(a)---means that for a fixed $z$ it is a function of $x^2+y^2$,
so one can set $y=0$. In addition, with $y=0$, $\Dl(\BC_g,\rho(\rB))$
for a fixed $x^2+z^2$ is monotone increasing in $z$ for $p \leq 1/2$, and
monotone decreasing for $p \geq 1/2$. Thus one can also set $x=0$ and look for
its maximum or minimum as a function of the single parameter $-1\leq z \leq 1$.
 
The range of the two parameters $p$ and $\lm$ for which $Q^{(1)}(\BC_g)$ is
greater than $0$ can be determined as follows. For $1/2 \leq p \leq 1$, $\BC_1$
is antidegradable~\cite{WolfPerezGarcia07}, while for $1/2 \leq \lm \leq 1$,
$\EC^\lm$ is antidegradable. Thus $\BC_g$, the concantenation of these two
channels (see \eqref{reversal} and the following discussion) is
antidegradable, and $Q^{(1)}(\BC_g)= Q(\BC_g)=0$. Thus $Q^{(1)}(\BC_g)$
can only be positive when both $p$ and $\lm$ are less than $1/2$.
At $p=0$, $\BC_1$ is a perfect channel and $\BC_g$ an erasure channel, so
$Q^{(1)}(\BC_g) = 1-2\lm$ for $\lm \leq 1/2$ (see \eqref{capEq} with $d_a =
2$). For $\lm=0$, $\BC_g$ is just the amplitude damping channel, which is
degradable with a positive $Q^{(1)}$ for $0\leq p <
1/2$~(see~\cite{WolfPerezGarcia07} and Sec.IV A in~\cite{SiddhuGriffiths16}).

For other values of $\lm$ and $p$ between $0$ and $1/2$, the numerical
maximization\footnote{Here and elsewhere in this work, numerical optimizations
use standard methods from SciPy~\cite{VirtanenGommersEA20}} of
$\Dl(\BC_g,\rho(\rB))$ together with an asymptotic analysis as $z$ approaches
$1$ ~(App.~\ref{Aq1GlnEra}) shows that $Q^{(1)}(\BC_g)$ is positive for $\lm$
in the interval
\begin{equation}
    0 \leq \lm  < \lm_0(p) = (1-2p)/(2-2p),
    \label{QBPos}
\end{equation}
(see Fig.~\ref{FigY}), is zero for $\lm \geq \lm_0(p)$, and as
\begin{equation}
 \dl\lm = \lm_0(p) - \lm
\label{dellm}
\end{equation}
tends to zero has the asymptotic form
\begin{equation}
 Q^{(1)}(\BC_g) \simeq a(p) \dl\lm \exp[ -b(p)/ \dl\lm],
\label{asm}
\end{equation}
where $a(p)$ and $b(p)$ are positive functions of $p$. The exponentially rapid
decrease of $Q^{(1)}(\BC_g)$ due to $\dl\lm$ in the denominator of the exponent
makes a direct numerical study difficult when $\dl\lm$ is very small. For
$p=1/4$ and $5 \times 10^{-3} < \dl\lm < 10^{-1}$ we find good agreement
between our numerical values and \eqref{asm}.

\begin{figure*}[ht]
  \centering
    \includegraphics[scale=.75]{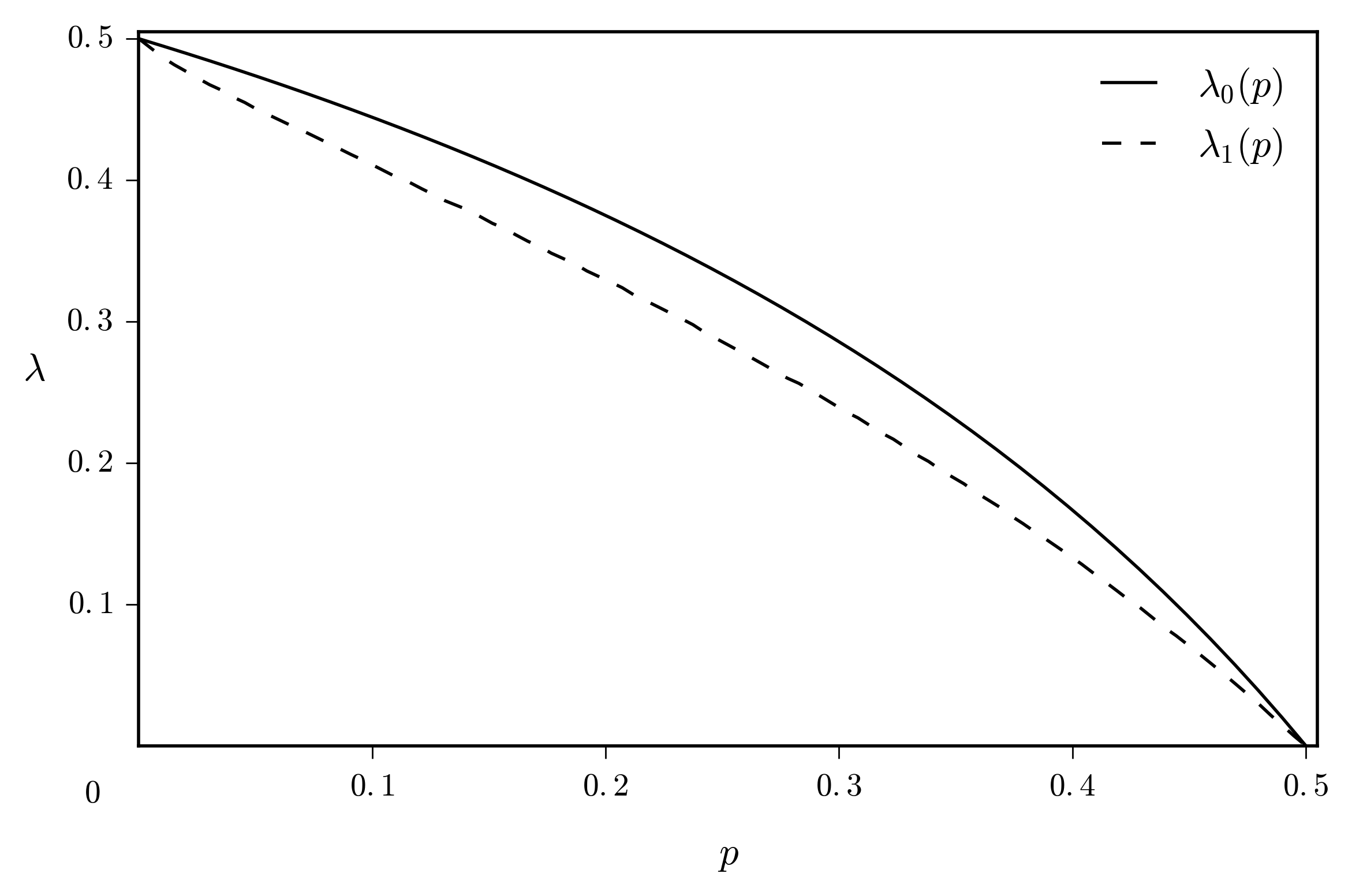}
 \caption{For a given $p$, $Q^{(1)}(\BC_g)$ is zero for $\lm \geq \lm_0(p)$,
and positive for $\lm < \lm_0(p)$. It is nonadditive at the 2-letter level
for $\lm_1(p) < \lm < \lm_0(p)$.}
    \label{FigY}
\end{figure*}

For all strictly positive $p$ and $\lm$, $Q^{(1)}(\CC_g)$ is positive, see
Sec. \ref{SEra}, so its complementary channel $\BC_g$ cannot be degradable, and
hence $Q^{(1)}(\BC_g)$ might conceivably be nonadditive. A relevant measure
of nonadditivity for $n$ copies of this channel placed in parallel is
\begin{equation}
    \dl_n := Q^{(1)}(\BC_g^{\ot n})/n - Q^{(1)}(\BC_g).
    \label{nonAddRes}
\end{equation}
One says that nonadditivity occurs at the \emph{$n$-letter level} if $n$
is the smallest integer for which $\dl_n > 0$.
We have found numerical evidence for nonadditivity at the 2-letter level for
$\lm$ in the range
\begin{equation}
 \lm_1(p) < \lm < \lm_0(p),\text{ thus } 
0 < \dl \lm <\dl\lm_1(p):=\lm_0(p) -\lm_1(p),
\label{eqn}
\end{equation}
where $\lm_1(p)$, determined numerically, is shown in Fig.~\ref{FigY}. In
particular for $\dl \lm$ between $0$ and $\dl\lm_1(p)$, an input density
operator for $\BC_g^{\ot2}$ of the form
\begin{equation}
 \sg = (1-\ep) [00] +\ep[\phi],\quad \ket{\phi} = (\ket{01} + \ket{10})/\st
\label{eqnsg}
\end{equation}
with $0<\ep<1$ chosen to maximize $\Dl(\BC_g^{\ot 2}, \sg)$
gives a larger maximum value of $\Dl(\BC_g^{\ot2})$ than the product
density operator 
\begin{equation}
 \tau = \rho_m\ot\rho_m, \quad \rho_m = (1-z)[0] + z [1],
\label{eqntau}
\end{equation}
with $0<z<1$ chosen to maximize $\Dl(\BC_g, \rho(\rB))$ for a single channel.
When $\dl\lm$ is sufficiently small, the asymptotic behavior of $\dl_2$
(App.~\ref{Aq1GlnEra}) is of the form \eqref{asm}, but with different choices
for $a(p)$ and $b(p)$. For larger $\dl\lm$, see Fig.~\ref{FigZ} for $p=0.25$,
it rises to a maximum and then falls to zero with a finite slope at
$\dl\lm=\dl\lm_1(p)$. While we can be quite confident of nonadditivity in the
region $\lm_1(p) < \lm < \lm_0(p)$, that $\dl_2$ is actually zero outside
this range is less certain, since an input density operator different from
\eqref{eqnsg} could conceivably give a maximum value of $\Dl(\BC_g^{\ot 2})$
larger than $2 Q^{(1)}(\BC_g)$, even though we have found no indication of this
in our numerical studies.

\begin{figure*}[ht]
  \centering
    \includegraphics[scale=.75]{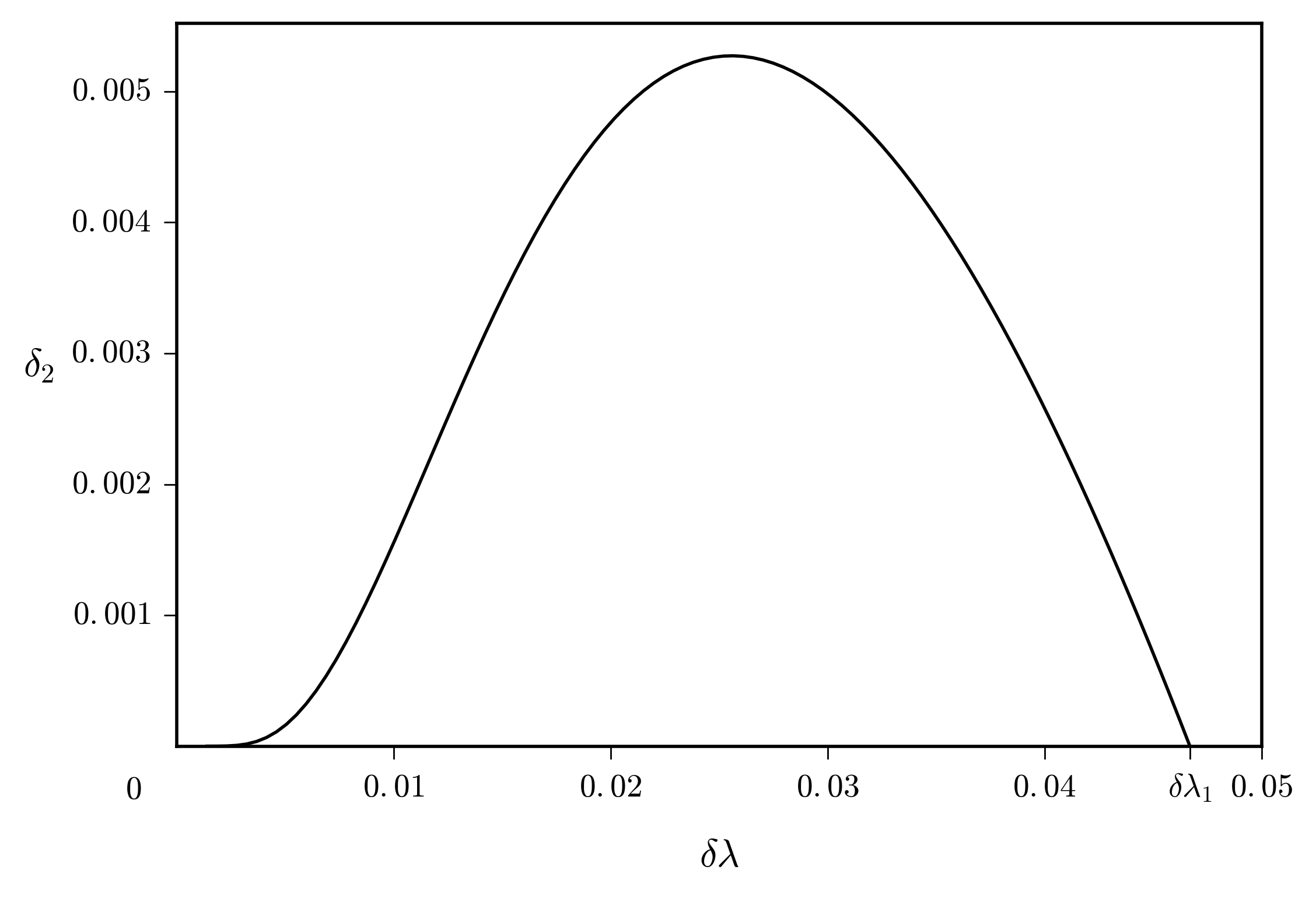}
    \caption{A plot of $\dl_2 = Q^{(1)}(\BC_g^{\ot 2})/2 - Q^{(1)}(\BC_g)$
    versus $\dl \lambda = \lm_0(p) - \lm$ at $p=0.25$ shows $\dl_2$ is positive
    for $0 <\dl\lm < 0.0406$ and attains a maximum value 
    $\simeq 5.27 \times 10^{-3}$.}
    \label{FigZ}
\end{figure*}

If $\dl_2$ is positive it is easy to show that $\dl_n$ is positive for all
$n>2$, and cannot be much smaller than $\dl_2$. As direct numerical searches
become exponentially more difficult with increasing $n$, it is customary to
make a guess or \emph{ansatz} $\rho_n$ for the input density operator, which
may depend upon a small number of parameters, and maximize
$\Dl(\BC_g^{\ot n},\rho_n)$ over these parameters, see \eqref{q1Def}, to obtain
a lower bound for $Q^{(1)}(\BC_g^{\ot n})$. When $n$ is even the \emph{pair
  ansatz} consists in dividing the $n$ channels into $n/2$ pairs and employing
the optimizing density operator $\sg$ defined above as the input for each pair;
this yields a lower bound $\dl_2$ for $\dl_n$. When $n$ is odd use $\sg$ for
each of $(n-1)/2$ pairs, and for the remaining channel the density operator
that gives rise to $Q^{(1)}(\BC_g)$; the resulting lower bound is a bit less
than $\dl_2$.
In the literature~\cite{DiVincenzoShorEA98, SmithSmolin07,
FernWhaley08,LeditzkyLeungEA18, BauschLeditzky20} various other ansatzes have been proposed,
including the $Z$-diagonal ansatz, a particular case of which is the repetition
ansatz. Our numerical studies for $n=3,4$ and 5 using these and others
motivated by the functional form of $\sg$ have not found any that improve on
the pair ansatz.

\subsection{Generalized Erasure with Qubit Dephasing Channel}
\label{Sdephasing}

The isometry $J_1 : \HC_{a1} \ra \HC_{b1}\ot \HC_{c1}$, with each space a qubit
(dimension 2), giving rise to the
dephasing channel $\BC_1$ and its complement $\CC_1$ can be 
written in the form
\begin{equation}
    J_1 \ket{0}_a = \ket{0}_{b1}\ket{\phi_0}_{c1}, \quad
    J_1 \ket{1}_a = \ket{1}_{b1}\ket{\phi_1}_{c1},
    \label{depIso}
\end{equation}
where
\begin{align}
    \nonumber
    \ket{\phi_0}_{c1} &= \sqrt{1-p}\ket{+}_{c1} + \sqrt{p}\ket{-}_{c1}, \\
    \nonumber
    \ket{\phi_1}_{c1} &= \sqrt{1-p}\ket{+}_{c1} - \sqrt{p}\ket{-}_{c1}; \\
    \ket{\pm} &:= (\ket{0} \pm \ket{1})/\sqrt{2},
\label{defphi}
\end{align}
with  $0\leq p\leq 1$ the dephasing probability. 
Interchanging $p$ with $1-p$ in \eqref{defphi} is equivalent to applying the
unitary $\sg_{z} = [0] - [1]$ to both $\HC_{b1}$ and $\HC_{c1}$, so for our
purposes we can limit $p$ to the range $0 \leq p \leq 1/2$.
The superoperators for the dephasing channel $\BC_1$ and its complement $\CC_1$ are 
\begin{align}
    \nonumber
    \BC_1(A) &= p Z A Z\ad + (1-p) A, \\
    \CC_1(A) &= \mte{0}{A}\, [\phi_0] + \mte{1}{A}\, [\phi_1],
    \label{dephasing}
\end{align}
where $Z \ket{0}_{a1} = \ket{0}_{b1}$ and $Z \ket{1}_{a1} = -\ket{1}_{b1}$. One
can think of $\CC_1$ as first measuring the input in the $\{\ket{0}, \ket{1}\}$
basis, and for measurement outcome $i$ preparing the channel output $[\phi_i]$.
Such a {\em measure-and-prepare} or {\em entanglement breaking} channel is
antidegradable and thus has zero quantum
capacity~\cite{HorodeckiShorEA03}.
The channel $\BC_1$ and its complement $\CC_1$ map 
$\rho(\rB)$ in \eqref{qBitBloch} to qubit density operators with Bloch vectors
\begin{align}
    \nonumber
    \rB_b &= ((1-2p)x,(1-2p)y,z), \quad \text{and} \\
    \rB_c &= (1-2p,0,2 \sqrt{p(1-p)} \; z),
    \label{dephDef}
\end{align}
respectively~(see Fig.~\ref{FigX}(b)).

Inserting $\BC_1$ and $\CC_1$ defined in \eqref{dephasing} in \eqref{glnEra1}
and \eqref{glnEra2} yields the generalized erasure channel pair $(\BC_g,
\CC_g)$. 
The channel $\BC_g$ is the same as the dephrasure channel studied
in~\cite{LeditzkyLeungEA18}, where it was defined using the second equality in
\eqref{reversal}. These authors showed that the global maximum or
minimum of $\Dl(\BC_g, \rho(\rB))$ occurs along $\rB = (x,0,z)$. They also
found that  for any fixed $p$ between $0$ and $1/2$, as $\lm$ increases from
$0$ a local maximum of $\Dl(\BC_g, \rho(\rB))$ remains at $x=z=0$
until $\lm$ reaches the value 
\begin{equation}
    j(p) = \frac{1 - 2p - 2p(1-p)\ln[(1-p)/p]}{2 - 4p - 2p(1-p)\ln [(1-p)/p]},
    \label{jp}
\end{equation}
at which point this maximum begins moving to positive $z$ values, while $x$
remains at $0$.
As $\lm$ increases further, the local maximum of $\Dl(\BC_g, \rho(\rB))$ goes
to zero at $\lm$ equal to
\begin{equation}
    g(p) = \frac{(1-2p)^2}{1 + (1 - 2p)^2},
    \label{gp}
\end{equation}
and remains zero for $\lm > g(p)$.

We strengthen these results by showing that for any $p$ and any $\lm$ between
$0$ and $1/2$ the \emph{global} maximum $Q^{(1)}(\BC_g)$ of
$\Dl(\BC_g, \rho(\rB))$ occurs along $\rB = (0,0,z)$, where it agrees with the
local maximum found in~\cite{LeditzkyLeungEA18}, while the global minimum,
equal to $-\,Q^{(1)}(\CC_g)$~(see \eqref{q1Def}) occurs on the line
$\rB = (x,0,0)$. 
To show these results, we first use rotational symmetry, see Fig.~\ref{FigX}.
This symmetry ensures $\Dl(\BC_g, \rho(\rB))$ only depends on $x^2 + y^2$.  The
dependence guarantees a maximum or minimum of $\Dl(\BC_g, \rho(\rB))$ can
always be obtained at $y=0$. Setting $y=0$ makes $\Dl(\BC_g, \rho(\rB))$ a
function of two independent variables $x$ and $z$ where $x^2 + z^2 \leq 1$.  To
understand $\Dl(\BC_g, \rho(\rB))$ better, we switch to different independent
variables $u = \sqrt{x^2 + z^2}$ and $v = z$, where $-u \leq v \leq u$ and $u
\leq 1$.  In these variables $\Dl(\BC_g, \rho(\rB))$ is a convex function of
$v$ for any fixed $u$; in addition, this convex function is symmetric about $v
= 0$.  Together the convexity and symmetry ensure that $\Dl(\BC_g, \rho(\rB))$
is maximum at $v = \pm u$ and minimum at $v = 0$. Translating these minimum and
maximum conditions in the $u$ and $v$ variable to the $x$ and $z$ variables shows that
the global maximum of $\Dl(\BC_g, \rho(\rB))$ occurs along $\rB = (0,0,z)$, and
its global minimum along $\rB = (x,0,0)$.

The study in~\cite{LeditzkyLeungEA18} used a repetition ansatz,
\begin{equation}
    \hat \rho_2(\eta) =  \eta [00] + (1-\eta) [11],
     \label{repAnz}
\end{equation}
with $\eta$ chosen appropriately between zero and one to maximize
$\Dl(\BC_g^{\ot 2},\hat\rho_2(\eta))$, and showed that $Q^{(1)}(\BC_g)$ is
non-additive at the two-letter level. Recently another
study~\cite{BauschLeditzky20}, used a neural network ansatz to obtain
non-additivity of $Q^{(1)}(\BC_g)$. In both studies, non-additivity is found
for some values of $p$ between $0$ and $1/2$, and some values of $\lm$ in the
range $j(p) < \lm < g(p)$.  
We extend these results by showing $\QC^{(1)}(\BC_g)$ is non-additive at
the two-letter level for all $0 < p < 1/2$ at $\lm = j(p)$. This extension is
obtained by using a different ansatz
\begin{equation}
  \rho(\zt)= \{ (1+\zt) \big([00] + [11]\big) + (1-\zt)\big([01]+[10]\big)\}/4,
     \label{zAnz}
\end{equation}
and varying $\zt$ between $-1$ and $+1$ to maximize $\Dl(\BC_g^{\ot
2},\rho(\zt))$. Inserting this maximum, $\Dl^*(\BC_g^{\ot 2})$, in
\eqref{nonAddRes} gives a lower bound $ \dl^*_2$ for $\dl_2$. We find that
along the curve $\lm = j(p)$, $0 < p <1/2$, $ \dl^*_2$ is positive and goes to
zero at the two end points~(see Fig.~\ref{FigW}). We also noticed that for
a fixed $p$, $ \dl^*_2$ rapidly goes to zero as $\lm$ increases or decreases
from $j(p)$. It remains an open question whether using a different ansatz than
\eqref{zAnz}, or by some other method, the range of $p$ and $\lm$ values for
which $Q^{(1)}(\BC_g)$ is nonadditive can be extended beyond those discussed
here and in previous studies~\cite{LeditzkyLeungEA18, BauschLeditzky20}.

\begin{figure*}[ht]
  \centering
    \includegraphics[scale=.75]{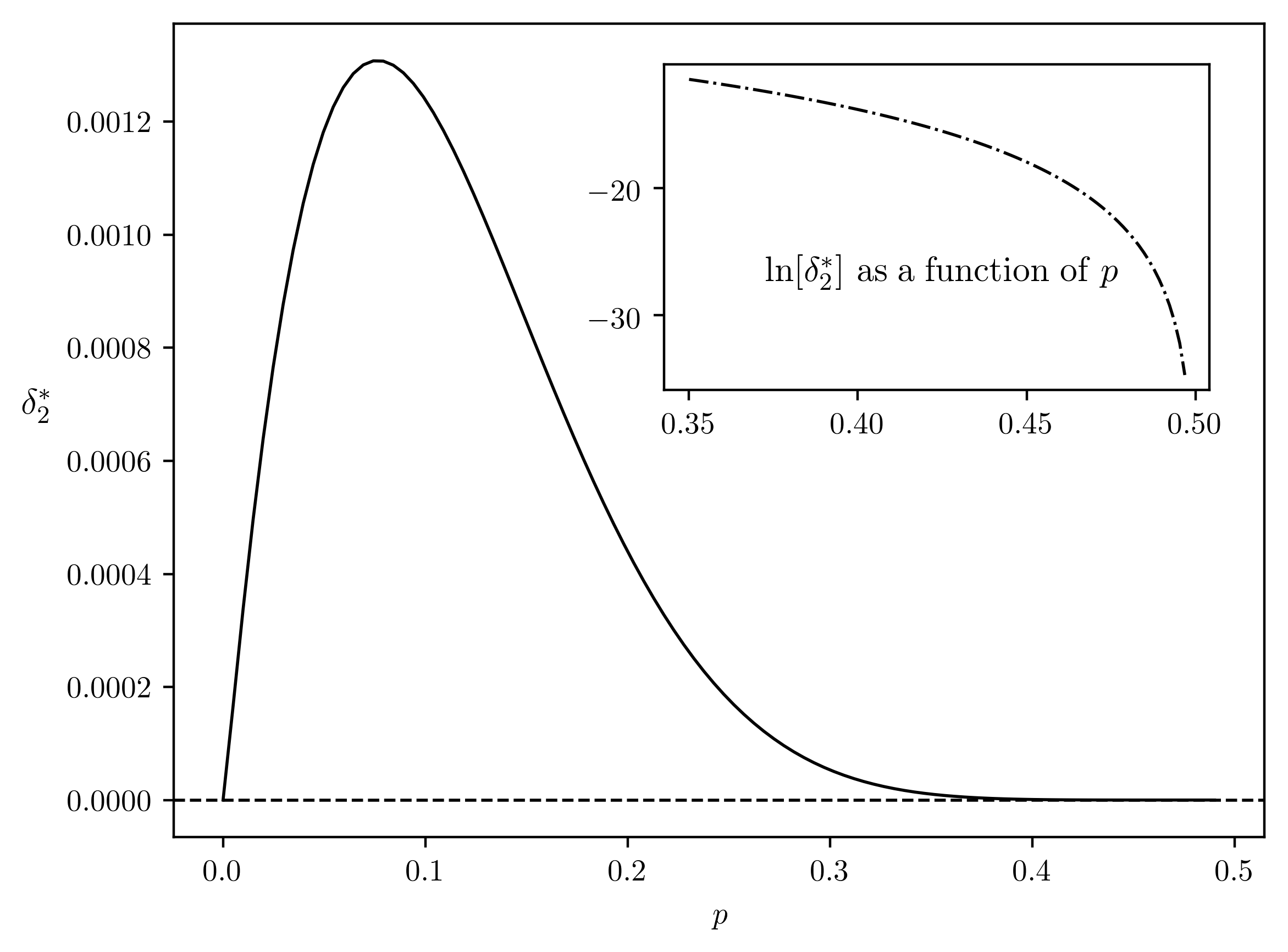}
    \caption{Plot of $\dl_2^* = \Dl^*(\BC_g^{\ot 2})/2 - Q^{(1)}(\BC_g)$ as a
    function of $p$ for the dephrasure channel $\BC_g$, with $\lm=j(p)$.}
    \label{FigW}
\end{figure*}

\subsection{Incomplete Erasure Channel}
\label{SEra}

\begin{figure*}[ht]
  \centering
    \includegraphics[scale=.75]{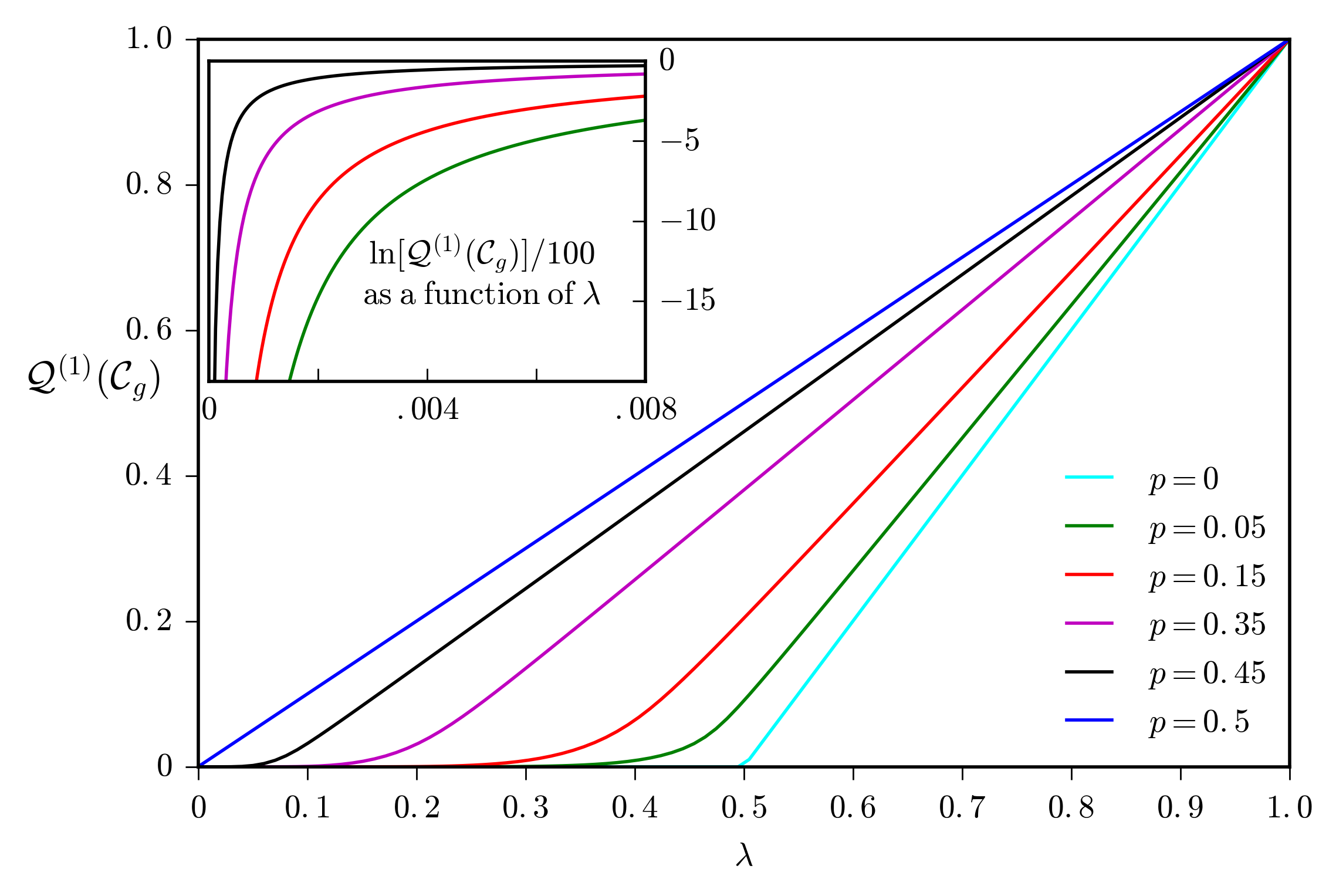}
    \caption{Plot of $Q^{(1)}(\CC_g)$ against $\lm \in [0,1]$ for various $p$
      values. The inset shows $Q^{(1)}(\CC_g)$ on a logarithmic scale for small
      positive $\lm$ and $p$.}
    \label{Q1C}
\end{figure*}

As discussed following \eqref{glnEra2}, $\CC_g$ resembles an erasure channel
$\CC_e$, except that instead of completely erasing its input it sends it
through a noisy channel $\CC_1$. This ``incomplete erasure'' leads to an
interesting effect shown in Fig.~\ref{Q1C} for the amplitude damping case of
Sec.~\ref{SAmpDamp}. When $p=0$, which means that $\CC_1$ is the completely
noisy trace channel $\TC$, both $Q^{(1)}(\CC_g)$ and $Q(\CC_g)$ are exactly
zero for $0\leq \lm \leq 1/2$. But as soon as $p$ is positive by the smallest
amount, $Q^{(1)}(\CC_g)$ is positive over the entire range $\lm > 0$. As $p$
tends to $0$, the analysis in App.~\ref{Aq1GlnEra} yields the asymptotic
behavior
\begin{equation}
  Q^{(1)}(\CC_g) \simeq \frac{\lm}{\ln 2} \exp[ (p + \ln p)(1-\lm)/\lm ] 
    \label{qCAsm}
\end{equation}
for $0< \lm \leq 1/2$; see the inset in Fig.~\ref{Q1C}. 
A very similar behavior is found when $\CC_1$ is the
complement of the phase-damping channel $\BC_1$ discussed in
Sec.~\ref{Sdephasing}, for which the corresponding asymptotic expression is 
\begin{equation}
    Q^{(1)}(\CC_g) \simeq \frac{\lm}{\ln 2} 
  \exp\big\{ [p + (1-2p)\ln p](1-\lm)/\lm \big\}.
    \label{qCAdp}
\end{equation}

In both cases, when $p$ is small $\CC_1$ is not only noisy but
antidegradable, so that its quantum capacity is exactly zero. Thus its ability
to make $Q^{(1)}(\CC_g)$ positive in the entire range $0 < \lm \leq 1/2$ comes
as something of a surprise.
These two examples might suggest that $Q^{(1)}(\CC_g)$ is positive for all
$\lm >0$ if $\CC_1$ is \emph{any} channel that is not completely noisy. But
this is not the case. If $\CC_1$ is an erasure channel with erasure probability
$\mu$, $\CC_g$ is an erasure channel with erasure probability  
\begin{equation}
 \ep = (1-\lm)\mu,
\label{erase}
\end{equation}
and thus has zero capacity for $\ep \geq 1/2$. This means $Q^{(1)}(\CC_g)=0$
for
\begin{equation}
 0\leq\lm\leq 1-1/(2\mu),
\label{eraseb}
\end{equation}
which is a finite interval for any $\mu$ greater than $1/2$. One way to derive
\eqref{erase} is to note that if one defines the transmission probability of an
erasure channel as $1$ minus the erasure probability, $\BC_1$ has a
transmission probability of $\mu$. Since $\BC_g$ is the concatenation
\eqref{reversal} of an erasure channel with transmission probability $\mu$ with
one whose transmission probability is $1-\lm$, it is an
erasure channel with transmission probability equal to the product $\mu(1-\lm)$.
And this is the erasure probability of its complement $\CC_g$.

\section{Summary and Conclusions}
\label{Scon}

Following a review in Sec.~\ref{Sprelim} of how an isometry gives rise to 
quantum channel pair $(\BC,\CC)$, the process of combining channels or channel
pairs that we call gluing is presented in Sec.~\ref{Sgluing}. 
Combining channels by placing them in parallel or series is of course 
well known. However, the gluing procedure, aside from particular cases
like convex combinations and direct sums of channels, has so far as we know 
not been discussed earlier in the literature, and might well have some 
interesting applications in addition to those discussed in this paper.

Our focus is on a type of gluing procedure that leads to what we call a
\emph{block diagonal} channel pair, see \eqref{glueBC} and the discussion
following it. What makes this procedure of combining channels particularly
useful for the study of quantum channel capacities is that the entropy of the
output of a block diagonal channel is a weighted sum of the entropies of the
outputs of the individual channels in the combination, plus a ``classical''
term, \eqref{concEntB}. Thus the entropy bias or coherent information, the
difference of the entropies of the outputs of a block diagonal channel pair
$\BC$ and $\CC$ for a given input, is a similar weighted sum (with the
``classical'' term cancelling out), \eqref{concEntBias}. In addition one has a
simple physical picture: with some probability the input to a block diagonal
channel pair is sent into one of several different channel pairs.

The quantum \emph{erasure channel} with erasure probability $\lm$, together
with its complement, an erasure channel with erasure probability $1 - \lm$, is
an example of a block diagonal channel pair formed by gluing together two
perfect channel pairs, as discussed in Sec.~\ref{SglnEra}. Replacing one of the
perfect channel pairs with an arbitrary pair $(\BC_1,\CC_1)$ results in a
\emph{generalized erasure} channel pair $(\BC_g, \CC_g)$. The channel $\BC_g$
can be viewed as a concatenation of $\BC_1$ with a suitable erasure channel,
while one can think of $\CC_g$ as an erasure channel with incomplete erasure.

In Sec.~\ref{SApp} we have analyzed two cases of generalized erasure channel
pairs constructed using a pair $(\BC_1,\CC_1)$, with both $\BC_1$ and $\CC_1$
qubit-to-qubit channels. In the first case, Sec.~\ref{SAmpDamp}, $\BC_1$ and
$\CC_1$ are complementary amplitude damping channels. In the second case,
Sec.~\ref{Sdephasing}, $\BC_1$ is a phase-damping channel, with complement
$\CC_1$ a measure-and-prepare channel. This second case has been studied
in~\cite{LeditzkyLeungEA18}; some of our results confirm and extend the ones
published there. In both cases the qubit channel pair depends on a single
parameter $0\leq p\leq 1$, which determines the amount of amplitude or phase
damping of $\BC_1$. Hence the corresponding generalized erasure channel is
characterized by two parameters: $p$ and the erasure probability $\lm$. Both
the amplitude and phase damping cases exhibit interesting, and to some extent
unexpected, behavior.

The nonadditive behavior of $Q^{(1)}(\BC_g)$ for two identical channels in
parallel is analyzed in Sec.~\ref{SAmpDamp} for the phase-damping case,
starting with a global optimization, assisted by an asymptotic analysis, to
yield accurate values of $Q^{(1)}(\BC_g)$. In the $(p,\lm)$ plane
$Q^{(1)}(\BC_g)$ is zero for $\lm \geq \lm_0(p)$, the upper curve in
Fig.~\ref{FigY}, and positive for $0 \leq \lm < \lm_0(p)$. A numerical search
for a positive $\dl_2 = Q^{(1)}(\BC_g^{\ot 2})/2-Q^{(1)}(\BC_g)$ suggests a
plausible form \eqref{eqnsg} for the bipartite input density operator, and
using this one finds a well-defined region in the $(p,\lm)$ plane, lying
between the two curves in Fig.~\ref{FigY}, in which $\dl_2$ is positive.
Figure~\ref{FigZ} shows $\dl_2$ as a function of $\dl\lm = \lm_0(p)-\lm$ for
$p=1/4$.

The nonadditivity of $Q^{(1)}(\BC_g)$ for the phase-damping case (the
dephrasure channel) was first studied in~\cite{LeditzkyLeungEA18}. Our global
optimization results confirm their $Q^{(1)}(\BC_g)$ calculation, showing that
it is zero for $\lm \geq g(p)$ and positive for $\lm < g(p)$, and we have
extended the range of $(p,\lm)$ values over which nonadditivity occurs, without
determining its full extent. See the discussion following~\eqref{gp} in
Sec.~\ref{Sdephasing}, and Fig.~\ref{FigW}.

Even when nonadditivity is absent for two identical channels in parallel it
could be present for three or more. One very preliminary result in
Sec.~\ref{SAmpDamp} for the multiple channel case suggests an interesting
possibility: There might be channels for which $Q^{(1)}$ is nonadditive when
two are placed in parallel, but thereafter no additional nonadditivity arises
when a collection of such ``double'' channels are placed in parallel with one
another.

The behavior of $Q^{(1)}$ of the incomplete erasure channel $\CC_g$, discussed
in Sec.~\ref{SEra}, is also quite surprising. In both the amplitude and phase
damping cases, when $p = 0$ the channel $\CC_g$ becomes an erasure channel with
erasure probability $1-\lm$, so that $Q^{(1)}(\CC_g) = Q(\CC_g) = 0$ for
$\lm \leq 1/2$. However, as soon as $p$ is positive, $\CC_g$ is greater than
zero over the range $0< \lm \leq 1$, see Fig.~\ref{Q1C}. It is surprising that
``assisting'' the erasure channel with a very noisy $\CC_1$, which itself has
zero quantum capacity, gives rise to this effect. Not every noisy $\CC_1$
provides such a dramatic improvement, and it would be interesting to determine
which channels do so. While the behaviour of $Q^{(1)}(\CC_g)$ in Fig.~\ref{Q1C}
emerges quite clearly from the mathematics, we lack an intuitive explanation.

The results reported here could be extended in various ways. Two real numbers,
$p$ and $q$, are needed to parametrize the family of channel pairs
$(\BC_1,\CC_1)$ where both are qubit-to-qubit channels. The amplitude- and
phase-damping cases discussed above correspond to different choices of $q$. It
may be possible to extend the results in Sec.~\ref{SApp} to this larger family
of channels; however the absence of certain symmetries that simplified the
analysis in Sec.~\ref{SApp} might lead to complications.

In addition, nonadditivity can, and undoubtedly does, occur in certain cases
when two \emph{nonidentical} $\BC_g$ channels, with unequal choices for the
parameter pair $(\lm,p)$, are placed in parallel. We have no idea what might
arise from a study of these, but analyzing what happens when the parameters
$(\lm,p)$ for one channel are varied while those for the other are held fixed
might in some situations turn out be simpler than studying identical channels
in which the two sets of parameters are identical.

The main advantage of the generalized erasure approach for studying positivity
and nonadditivity of $Q^{(1)}$ lies in the fact that when two channel pairs
with a very simple structure, in our case the $(\BC_1,\CC_1)$ pair and the
perfect channel pair, are glued together, this can give rise to new and
interesting behavior not present in either of the separate components. One
suspects that there are other instances of this sort worth exploring, and one
can hope that analyzing them will yield additional insights into the behavior
of the quantum capacity of noisy quantum channels---a very challenging, but at
the same time very important, problem in quantum information theory, something
which needs to be better understood. We hope our results, limited as they are,
will make some contribution to this end.

\section*{Acknowledgments}
We thank Mark M. Wilde, Felix Leditzky, and two anonymous referees for helpful
comments. This work used the Extreme Science and Engineering Discovery
Environment (XSEDE)~\cite{TownsCockerillEA14}, which is supported by National
Science Foundation grant number ACI-1548562. Specifically, it used the Bridges
system~\cite{NystromLevineEA15}, which is supported by NSF award number
ACI-1445606, at the Pittsburgh Supercomputing Center (PSC).

\appendices

\section{Appendix. Concatenation and antidegradable channels}
\label{Aad}

Figure~\ref{antiMaps} will be of assistance in understanding the following proof that the
concatenation 
\begin{equation}
    \BC = \BC_2 \circ \BC_1,
    \label{compMap}
\end{equation}
of two channels placed in series is
antegradable if either $\BC_1$ or $\BC_2$ is antidegradable. Let
\begin{equation}
 J_1: \HC_a \ra \HC_{b1}\ot \HC_{c1}, \quad 
J_2: \HC_{b1} \ra \HC_{b2}\ot \HC_{c2}, 
\label{eqna2}
\end{equation}
be the isometries that give rise to the channel pairs $(\BC_1,\CC_1)$ and
$(\BC_2,\CC_2)$. Then
 $\BC$ and its complement $\CC$ are generated by the isometry $J:\HC_a
\ra\HC_b\ot\HC_c$, where
\begin{equation}
 \HC_b = \HC_{b2},\quad \HC_c = \HC_{c1}\ot\HC_{c2},\quad 
 J=(J_2\ot I_{c1})\circ J_1,
\label{eqna3}
\end{equation}
with $I_{c1}$ the identity map on $\HC_{c1}$. Thus the complement $\CC$ of
$\BC$ maps $\hat\HC_a$ to the tensor product $\hat \HC_{c1} \ot \hat\HC_{c2}$,
and its partial traces over these outputs are:
\begin{equation}
\Tr_{c1}[\CC(A)]=\CC_2\circ\BC_1(A),\quad \Tr_{c2}[\CC(A)] = \CC_1(A)
\label{eqna4}
\end{equation}

If $\BC_1$ is antidegradable, i.e., $\CC_1$ is degradable, there exists a
degrading map $\DC_1:\hat\HC_{c1} \ra \hat\HC_{b1}$, indicated by a
dashed curve in Fig.~\ref{antiMaps} (ignore $\DC_2$), such that for
any $A$ in $\hat\HC_a$,
\begin{equation}
 \DC_1\circ \CC_1(A) = \BC_1(A)
\label{eqna5}
\end{equation}
Given $\DC_1$, one can define a degrading map $\DC$ that maps
$\hat\HC_{c}= \hat\HC_{c1}\ot\hat\HC_{c2}$ to $\hat\HC_{b} = \hat\HC_{b2}$ by
its action on the tensor product of operators $F_1\in \hat\HC_{c1}$ and 
$F_2\in \hat\HC_{c2}$,
\begin{equation}
 \DC(F_1\ot F_2) =  \Tr(F_2)\cdot\BC_2\circ\DC_1(F_1).
\label{eqna6}
\end{equation}
Then use linearity to extend this to the entire operator space $\hat\HC_{c}$.
Intuitively (Fig.~\ref{antiMaps}) $\DC$ ``throws away'' the $\hat\HC_{c2}$ output, while
mapping the $\hat\HC_{c1}$ output to $\hat\HC_{b1}$. Since $\CC_1$ followed by
$\DC_1$ is the same as $\BC_1$, $\DC\circ\CC$ is identical to
$\BC=\BC_2\circ\BC_1$. Hence $\CC$ is degradable and $\BC$
antidegradable.

If instead of $\BC_1$ we assume that $\BC_2$ is antidegradable, the appropriate
degrading map $\DC:\hat\HC_c\ra\hat\HC_{b1}$ is obtained by ``throwing away''
the $\hat\HC_{c1}$ output of $\CC$ and applying the degrading may $\DC_2$, see
Fig.~\ref{antiMaps} and ignore $\DC_1$, to the $\hat\HC_{c2}$ output, with the result
\begin{equation}
 \DC(C_1\ot C_2) =\Tr(C_1)\cdot \DC_2(C_2).
\label{eqna7}
\end{equation}
So again $\DC\circ\CC$ is identical to $\BC=\BC_2\circ\BC_1$, which means $\CC$
is degradable and $\BC$ antidegradable.

\begin{figure}[ht]
    \centering
    \includegraphics[scale=1]{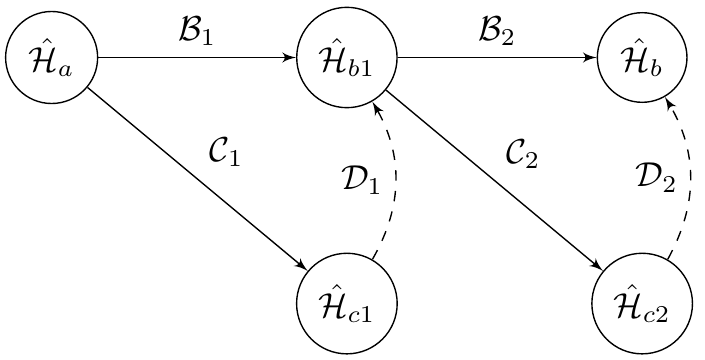}
    \caption{A schematic diagram indicating the spaces 
    $\hat \HC_a, \hat \HC_{b1}, \hat \HC_{b}, \hat \HC_{c1},$ and $\hat \HC_{c2}$,
    and the channels $\BC_1,\BC_2, \CC_1,\CC_2,\DC_1$, and $\DC_2$ acting
    between these spaces.}
    \label{antiMaps}
\end{figure}

\section{Appendix. Asymptotic estimates of $Q^{(1)}(\BC_g)$ and $Q^{(1)}(\CC_g)$}
\label{Aq1GlnEra}


Some use was made in Sec.~\ref{SApp} of asymptotic expressions for the 
channel coherent information $Q^{(1)}$ in circumstances in which a straightforward numerical
approach runs into difficulties because one is trying to find the maximum or
minimum of a function
\begin{equation}
 f(\ep) = \al\, \ep\ln(\ep) + \bt\,\ep,
\label{eqn1}
\end{equation}
where $\ep>0$ is small, and $\al$ and $\bt$ are real numbers. If $\al$ is
positive, $f(\ep)$ will be negative for sufficiently small $\ep$, and positive
if $\al$ is negative. If $\al$ and $\bt$ are both positive,  $f$ has
a minimum at
\begin{equation}
 \ep = \ep_m := \exp[-(1+\bt/\al)],
\label{eqn2}
\end{equation}
where it takes the value
\begin{equation}
  f(\ep_m) = -\al\,\ep_m = -\al \exp[-(1+\bt/\al)].
\label{eqn3}
\end{equation}
If both $\al$ and $\bt$ are negative, $f$ has a maximum rather than a minimum
at \eqref{eqn2}, and the maximum value is again given by \eqref{eqn3}.

A first application of these formulas is to the amplitude damping case,
Sec.~\ref{SAmpDamp}, where for $\rB=(0,0,z)$ in the expression for $\rho(\rB)$
in \eqref{qBitBloch}
\begin{equation}
f(\ep) = \Dl(\BC_g,\rho(\rB)),\quad \ep = 1-z
\label{eqn4}
\end{equation}
has the form \eqref{eqn1} for small $\ep$, where as a function of $0<\lm<1/2$
and $0<p<1/2$,
\begin{equation}
 \al = [p(1-\lm)+\lm-1/2]/\ln 2
\label{eqn5}
\end{equation}
This is zero along the line $\lm=\lm_0(p)$, \eqref{QBPos}, and negative when
$\dl \lm = \lm_0(p) - \lm$ is positive. Hence $\Dl(\BC_g,\rho(\rB))$, and
therefore its maximum $Q^{(1)}(\BC_g)$, is greater than zero for sufficiently
small $\dl\lm>0$. This is consistent with numerical results that indicate that
$Q^{(1)}(\BC_g)$ is zero for $\lm\geq \lm_0(p)$ and positive elsewhere.

One can work out the asymptotic form of $Q^{(1)}(\BC_g)$ for small positive
$\dl\lm$ using \eqref{eqn3} and
\begin{equation}
 \al = \al_1(p)\, \dl\lm,\quad 
 \bt = \bt_0(p) + \bt_1(p)\, \dl\lm 
\label{eqn6}
\end{equation}
where
\begin{align}
 \al_1(p) &= -(1-p)/\ln 2,
\notag\\
 \bt_0(p) &= [(p\ln p)/(1-p) -\ln(1-p)]/4 \ln 2,
\notag\\
 \bt_1(p) &= (1-p)[2\bt_0(p) + 1 + 1/\ln2],
\label{eqn7}
\end{align} 
Both $\al_1(p)$ and $\bt_0(p)$ are negative in the range of interest,
$0<p<1/2$, so $\Dl(\BC_g,\rho(\rB))$ will have a maximum at
\begin{equation}
\ep_m\simeq K \exp[ -\bt_0/(\al_1\, \dl\lm)],\quad
K =\exp[-1-(\bt_1/\al_1)],
\label{eqn8}
\end{equation}
and thus $Q^{(1)}(\BC_g)$, the maximum of $\Dl(\BC_g,\rho(\rB))$,
has the asymptotic form
\begin{equation}
 Q^{(1)}(\BC_g) \simeq -\al_1\,\dl\lm\, \ep_m \simeq
 a(p) \dl\lm\exp[-b(p)/\dl\lm],
\label{eqn9}
\end{equation}
for small $\dl\lm$, where 
\begin{equation}
 a(p) := -\al_1 K, \quad b(p) := \bt_0/\al_1,
 \label{eqn10}
\end{equation}
are positive functions whose $p$ dependence is determined by that of $\al$
and $\bt$. The final factor in \eqref{eqn9} is exponentially small due to the
$\dl\lm$ in the denominator of the exponent. The approximation \eqref{eqn9} is
in reasonable agreement with direct numerical calculations for small $\dl\lm$
at $p=1/4$.

In the same way one can find the asymptotic behavior, for $0 < \lm <1/2$ and
$p$ very small, of $Q^{(1)}(\CC_g)$, equal to \emph{minus} the minimum of
$f(\ep)=\Dl(\BC_g,\rho(\rB))$ in \eqref{eqn4}. In this case $z$ is close to
$-1$ and
\begin{equation}
 \ep = 1+z,
\label{eqn11}
\end{equation}
is a small quantity. Now $\al$ and $\bt$ are given by
\begin{align}
    \nonumber
    \al &= \lm/\ln 2, \\
    \bt &= \bt_0(\lm) +\bt_1(\lm)\ln p + \bt_2(\lm) p\ln p + \bt_3(\lm) p +\cdots,
\label{eqn12}
\end{align}
where 
%
\begin{align}
    \nonumber
    \bt_0 &= -\al (1+\ln 2),& 
    \bt_1(\lm) &=-\al (1-\lm)/\lm, &\\
    \bt_2(\lm) &= 0,  &
    \bt_3(\lm) &= \bt_1(\lm).&
\label{eqn13}
\end{align}
Inserting these in \eqref{eqn3} one arrives at the asymptotic formula
\eqref{qCAsm} for $Q^{(1)}(\CC_g) \simeq -f(\ep_m)$

An asymptotic estimate for small $\dl\lm$ of the nonadditity of
$Q^{(1)}(\BC_g)$ at the 2-letter level, see \eqref{nonAddRes}, can be carried
out assuming that $\ep$ in the input density operator $\sg$, \eqref{eqnsg},
is small, and looking for the maximum of
\begin{equation}
  \bar f(\ep) = \Dl(\BC_g^{\ot 2},\sg(\ep)) = \bar\al\ep\ln(\ep) +\bar\bt\ep.
\label{eqn14}
\end{equation}
It turns out that
\begin{equation}
 \bar\al=2\al,\quad
\bar\bt = \bar\bt_0(p) + \bar\bt_1(p)\,\dl\lm + \bar\bt_2(p)\,\dl\lm^2,
\label{eqn15}
\end{equation}
where $\al$ is the single channel quantity
defined in \eqref{eqn6}, and 
\begin{strip}
\begin{align}
 \bar\bt_0(p) &= \frac{(-2\ln2)p +(4\ln2)p^2 +p\ln p +(1-p(1+2p))\ln(1+p)
    -2(1-p)^2\ln(1-p)}{4\ln 2 (1-p)^2},
\notag\\ 
\bar\bt_1(p) &= \frac{2(1-2p)+2 p^2(1+\ln 2) + p \ln p -(1-p)^2\ln(1-p) 
-p(1+p)\ln(1+p)}{(1-p)\ln 2},
\notag\\
\bar\bt_2(p) &=[p\ln(4p) -(1+p)\ln(1+p)]/\ln 2.
\label{eqn16}
\end{align}
\hrulefill
\end{strip}
That $\bar\al=2\al$ makes it convenient to consider the ratio
\begin{equation}
 R = \frac{\bar Q^{(1)}(\BC_g^{\ot 2})}{2Q^{(1)}(\BC_g)} 
 = \frac{\bar \ep_m}{\ep_m} =\exp[(\bt/\al)(1-\bar\bt/2\bt)],
\label{eqn17}
\end{equation}
and thus
\begin{equation}
Q^{(1)}(\BC_g^{\ot 2}) \simeq \bar a(p) \dl\lm \exp[-\bar b(p)/\dl\lm],
\label{eqn17b}
\end{equation}
with 
\begin{equation}
 \bar a(p) = -2\al_1\exp[-1-\bar\bt_1/2\al_1],\quad
\bar b(p) =\bar\bt_0/2\al_1.
\label{eqn17c}
\end{equation}
Provided  
\begin{equation}
 \bar\bt_0/(2\bt_0) <1
\label{eqn18}
\end{equation}
a condition fulfilled for all $0 < p < 1/2$, $\bar b(p)$ is less than $b(p)$,
$R$ tends to $+\infty$ as $\dl\lm$ goes to zero, so that as $\dl\lm \ra 0$,
\begin{align}
    \dl_2 &= Q^{(1)}(\BC_g^{\ot 2})/2 -Q^{(1)}(\BC_g) \notag \\
    &\simeq  Q^{(1)}(\BC_g^{\ot 2})/2 \notag \\
    &\simeq \bar a(p) \dl\lm \exp[-\bar b(p)/\dl\lm],
\label{eqn18b}
\end{align}

In the case of the phase-damping channel, Sec.~\ref{Sdephasing}, similar
asymptotic estimates are possible, where the small parameter is now
\begin{equation}
 \dl\lm = g(p)-\lm,
\label{eqn19}
\end{equation}
where $g(p)$ is defined in \eqref{gp}.
For $Q^{(1)}(\BC_g)$ the 
coefficients in \eqref{eqn6} and \eqref{eqn7} are given by
\begin{align}
\al_1(p) &= -[1+(1-2p)^2]/2\ln2,
\notag\\
\bt_0(p) &= \frac{2p(1-p)\ln(4p(1-p))}{ (1+(1-2p)^2)\ln 2},
\notag\\
\bt_1(p) &= [1+\ln 2-2p(1-p)(1-\ln(2p(1-p))]/\ln 2.
\label{eqn20}
\end{align}
These coefficients when inserted in \eqref{eqn8} and \eqref{eqn10} yield
the asymptotic form \eqref{eqn9}.

Similarly, an asymptotic formula for $Q^{(1)}(\CC_g)$ is obtained by 
employing in \eqref{eqn12} the quantities
\begin{align}
    \al &=  \lm/\ln 2,&
    \bt_0 &= -\al (1+\ln 2),&\notag\\
    \bt_1(\lm) &=-\al (1-\lm)/\lm,&
    \bt_2(\lm) &= -2\bt_1(\lm),&\notag \\
    \bt_3(\lm) &= \bt_1(\lm)
\label{eqn21}
\end{align}

resulting in the asymptotic form \eqref{qCAdp}.


\end{document}